 \documentstyle[aas2pp4]{article}

\begin{document}

\title{Evidence for Rotation in the Galaxy at $z=3.15$ Responsible for
a Damped Lyman-alpha Absorption System in the Spectrum of Q2233+1310}
\author{Limin Lu$^{1}$, Wallace L. W. Sargent, \& Thomas A. Barlow}
\affil{California Institute of Technology, 105-24, Pasadena, CA 91125}
\altaffiltext{1}{Hubble Fellow}

\begin{abstract}

    Proof of the existence of a significant population of normal disk         
galaxies at redshift $z>2$ would have profound implications for
theories of structure formation and evolution. We present 
evidence based on Keck HIRES spectra
that the damped Ly$\alpha$ absorber at $z=3.15$ toward the
quasar Q 2233+1310 may well be such an example. Djorgovski et al.
have recently detected the Ly$\alpha$ emission from the absorber,
which we assume is at the systemic redshift of the absorbing galaxy. 
By examining the profiles of the metal absorption lines arising 
from the absorbing galaxy in relation to its systemic redshift, 
we find strong kinematical  evidence for rotation. 
Therefore the  absorber is likely to be a disk galaxy.
The inferred circular velocity for the galaxy is $\geq200$ km s$^{-1}$.
With a separation of $\simeq17$ kpc
($q_0=0.1$, $H_0=75$) between the galaxy and the quasar sightline, the
implied dynamic mass for the galaxy is $\geq 1.6\times 10^{11} M_{\odot}$.
The metallicity of
the galaxy is found to be [Fe/H]$=-1.4$, typical of damped Ly$\alpha$
galaxies at such redshifts. However, in another damped Ly$\alpha$
absorber at $z=2.81$ toward Q 0528$-$2505, no kinematical
evidence for galactic rotation is evident. In the latter case, the
damped Ly$\alpha$ absorber occurs near the background quasar in redshift
so its properties may be influenced by the background quasar.
These represent the only two cases at
present for which
the technique used here may be applied. Future applications of the
same technique to a large sample of damped Ly$\alpha$ galaxies may
allow us to determine if a significant population of disk galaxies already
existed only a few billion years after the Big Bang.

\end{abstract}


\section{INTRODUCTION}

   The existence (or lack) of a significant population of normal
disk galaxies at very high redshifts ($z>2$) would have profound
implications for models of structure formation and evolution in the
early universe.  While the damped
Ly$\alpha$ (DLA) systems seen in spectra of high-redshift quasars
have been suggested to represent disk galaxies
in their youth (Wolfe et al 1986), 
direct evidence for such a hypothesis has been scarce
(see the discussion by Lu et al 1996). Wolfe (1995) has suggested that
the metal absorption lines associated with DLA
galaxies exhibit what was termed ``edge-leading'' asymmetric profiles,
as expected from rotating gaseous structures.
Detailed analysis of a larger sample of DLA systems
appears to confirm this result (Prochaska \& Wolfe 1997).

   The DLA system at $z_{damp}=3.15$ in the spectrum of
the background quasar Q 2233+1310 ($z_{em}=3.30$) has a neutral hydrogen 
column density $N$(H I)$\simeq 10^{20}$ atoms cm$^{-2}$ (Lu et al 1993). 
Very recently, Djorgovski et al (1996) 
reported the detection of Ly$\alpha$ emission
at $z=3.1530$ only 2.3" away from the quasar. 
They argued that the Ly$\alpha$ emission most likely comes from the 
same galaxy that is also responsible for the DLA absorption
seen in the quasar spectrum. The separation between the
galaxy and the quasar sightline implies a radius of
at least 17 kpc ($q_0=0.1$ and $H_0=75$) for the galaxy,
comparable to the size of normal spiral disks. 
Djorgovski et al suggested that the offset of $\sim 200$ km s$^{-1}$
between the redshift of the Ly$\alpha$ emission and that of the damped
Ly$\alpha$ absorption may be due to galactic rotation, although no
independent supporting evidence was available.
 
   The detection of the Ly$\alpha$ emission provides a measure
of the systemic redshift
of the absorbing galaxy. An examination of the metal absorption line
profiles in relation to the systemic redshift may then reveal important
kinematic information about the absorbing gas and hence the nature of the
absorbing galaxy. 
Motivated by this, we obtained an echelle spectrum of 
Q 2233+1310 in July 1996 using the 10m Keck telescope and the
HIRES spectrograph.  Details of the observations,
data reductions, and full analysis of the DLA system
 will be described elsewhere (Lu, Sargent, \& Barlow 1997, in preparation). 
Here we only discuss the results relevant
to the galactic rotation hypothesis.
 
\section{ANALYSIS AND RESULTS}

  Figure 1 shows the profiles of selected low-ionization
metal absorption lines from the DLA absorber, as well as those
of the high-ionization C IV and Si IV lines. Djorgovski et al (1996)
gave a redshift for the Ly$\alpha$ emission of $z=3.1530\pm0.0003$,
which we adopt as the systemic redshift of the absorbing 
galaxy\footnote{Absorption by dust and resonant scattering by H atoms within
the absorbing galaxy can change significantly the energy distribution of the 
escaped Ly$\alpha$ photons and hence the profile of the Ly$\alpha$
emission. However, unless the processes are
very asymmetric with respect to the rest-frame velocity of the 
galaxy, the systemic redshift as inferred from the Ly$\alpha$
emission should be largely unaffected.}.
It is important to keep in mind that, 
although the measurement uncertainty in the redshift of the Ly$\alpha$
emission is only $\Delta z=0.0003$ or $\Delta v=22$ km s$^{-1}$,
the systematic error could be significantly larger because the blue wing
of the Ly$\alpha$ emission may be absorbed by Ly$\alpha$ forest clouds
at lower redshifts. Indeed, the blue side of the Ly$\alpha$ emission profile
appears much steeper than the red side 
(see figure 3 of Djorgovski et al 1996), making the
above a valid concern.  Hence the true systemic redshift of the absorbing
galaxy is probably somewhat less than the quoted value of $z=3.1510$.

\subsection{Evidence for Galactic Rotation}

  The main purpose of this study is to examine the kinematics of the 
absorbing gas as revealed by the metal absorption lines to see if there is
any evidence for galactic rotation. The idea behind
such an analysis was first discussed by Lanzetta \& Bowen (1992) and then
by Wolfe (1995) and Prochaska \& Wolfe (1997). Here we describe 
the basic ideas but refer the
readers to  Prochaska \& Wolfe (1997) for details.

Take the simplest
case where the quasar sightline passes through a galactic disk edge-on
but off the center of the galaxy.
The galaxy is assumed to have a finite thickness and 
a flat rotation curve beyond the innermost region.
Further assume that the distribution of the gas in the
disk is such that the density of the gas  decreases with
galactocentric radius as well as with the distance away from the
galactic midplane. Then the point of closest impact
to the galactic center along the quasar sightline should have the strongest
absorption, and also the largest projected velocity from the rotation.
Conversely, the points along the quasar sightline that are further away
from the galactic center will have less absorption and smaller projected
velocities (see figure 3 of Prochaska \& Wolfe 1997).
The result is that the strongest
absorption (which is associated with the closest 
point of impact in this example) should occur at
either the blue or the red edge of the absorption profile (depending
on the direction of the galactic rotation), and that the strength of the 
absorption should decrease smoothly toward the opposing edge.
Such profiles were termed ``edge-leading asymmetric'' by Prochaska \&
Wolfe (1997). More general considerations with realistic disk models 
and sightline orientations are discussed
by Prochaska \& Wolfe (1997), but the edge-leading asymmetry is generally
preserved.  Replacing the smooth gas distribution
with discrete cloud distribution does not change the qualitative picture,
although the random nature of the cloud distribution means that edge-leading
asymmetric profiles may not be apparent in every single case.

  In the case of Q 2233+1310, the metal absorption line profiles are clearly
asymmetric, with the strongest absorption component occurs near the blue
edge and with the strength of the components decreases smoothly toward the
red edge, as is best illustrated by the Fe II $\lambda$1608 profile in 
figure 1\footnote{Note that weak, unsaturated low-ionization
absorption lines should be the best tracer of disk gas. Strong, saturated
absorption lines (such as O I$\lambda$1302 shown in figure 1) are less
suited for this purpose because (1) the information on the velocity
structure of the absorption is often lost in the saturated parts,
and (2) strong absorption lines are more sensitive to trace amounts of
gas so they are more easily contaminated by diffuse clouds (e.g., halo
clouds) that are not associated with the galactic disk. Absorption lines
of high ionization species (e.g., Si IV, C IV) are not suited for this
purpose either because they are not expected to trace the bulk of neutral
gas in galactic disks.}.
These are characteristic of profiles caused by rotating disk-like structures.
Figure 3 of Djorgovski et al (1996) shows that the absorbing galaxy 
is clearly at a slightly
higher redshift than the damped Ly$\alpha$ absorption itself.
This means that the dominant H I absorbing gas along the quasar sightline
must be moving toward us. Under the galactic
rotation hypothesis, this means that the projected rotation velocity of
the galaxy along the quasar sightline must be pointing toward us.
Consequently, the metal line profiles in the Q 2233+1310 DLA
system can be understood if the absorbing galaxy is a rotating disk and
if the quasar line of sight intersects the midplane of the disk near the
major axis\footnote{The major axis here is defined as the intersection
between the plane of the disk and the plane of the sky. The sightline
must intersect the midplane of the disk near the major axis
because, if the sightline intersects the midplane of the disk far from the
major axis, a reverse asymmetry in the absorption line profile may occur
(ie, the strongest component will appear near the red edge rather than
the blue edge). An example of this is shown in case 1 of figure 3 of
Prochaska \& Wolfe 1997.}. The galactic rotation hypothesis then predicts 
that the velocity spread in the
absorption components should be comparable to or smaller than
the maximum projected rotation velocity
along the quasar sightline. For the Q 2233+1310 system, the components in
the Fe II absorption has a total velocity spread of 
$\Delta v \sim 200$ km s$^{-1}$,
which, due to projection effects,  should be a lower limit to the 
true rotation velocity. The galactic rotation hypothesis also predicts
that the strongest absorption component
should be at the velocity corresponding to the maximum projected velocity
along the quasar line of sight (relative to the systemic 
velocity of the galaxy). This would indicate that 
the maximum projected rotation velocity of the galaxy
is $\sim 270$ km s$^{-1}$ based on Fe II$\lambda$1608 in figure 1. 
Part of the difference between this velocity and the projected velocity 
of $\geq 200$ km s$^{-1}$ inferred from the velocity spread
of the absorption components may be due to a small systematic
error in the systemic redshift of the absorbing galaxy caused by absorption
from foreground Ly$\alpha$ forest clouds (see discussion at the beginning
of the section).

   In summary, the edge-leading asymmetry seen in the metal absorption
lines and the total velocity spread of the components in the Q 2233+1310
DLA system are entirely consistent
with the hypothesis (Djorgovski et al 1996)
that the absorbing galaxy is rotating with a 
circular velocity of $v_{rot}>200$ km s$^{-1}$. In fact, these 
pieces of evidence
may be taken as support for the galactic rotation hypothesis because, if 
the absorbing clouds were moving randomly, there is no {\it a priori} reason
why (1) the strongest absorption component should be at the blue edge of
the absorbing complex and why the strength of the absorption components
should decrease smoothly toward increasing velocity, (2) the total
velocity spread among the components should be similar
to that inferred from the position of the strongest absorption component.
Gas clouds moving in predominantly radial orbits are expected to produce
absorption profiles that are symmetric with respect to the systemic
velocity of the absorbing galaxy with absorption peaks
occurring at both the blue and red edges of the profiles
(Lanzetta \& Bowen 1992; Prochaska \& Wolfe 1997),
contrary to what is observed here.

     A potential problem with the galactic disk interpretation for the
Q 2233+1310 DLA system is the
large velocity spread of the absorbing components: $\Delta v\sim 200$
km s$^{-1}$. Monte Carlo simulations by Prochaska \& Wolfe (1997)
indicate that, for given rotation velocity and inclination angle,
$\Delta v$ decreases with increasing impact parameter. The
large observed $\Delta v$ for the Q 2233+1310 system requires
the absorbing galaxy to have a very large disk (with an
exponential disk scale
length comparable to the impact parameter$\simeq$17 kpc) and to be viewed
nearly edge-on. Spiral disks with such large scale length do exist
but are quite rare in the local universe (cf. Grosbol 1985;
Kent 1985; note the different choices of Hubble constant in these
studies). 
However, the technique of damped Ly$\alpha$ absorption
may preferentially pick out large galaxies owing to their large
absorption cross sections. The requirement on the scale length 
may be relaxed if some of the
velocity spread is due to non-circular motions (eg, random motions) 
of the clouds. In any case, it may be possible to test these predictions
with deep imaging observations.

   The width of the Ly$\alpha$ emission profile from the Q 2233+1310
DLA absorber provides a consistency check of the disk
interpretation. The Ly$\alpha$ emission line has an apparent
width of FWHM$\sim 360$ km s$^{-1}$ (see figure 3 of Djorgovski et al
1996), which, due to possible absorption of the blue wing by
Ly$\alpha$ forest clouds, is likely to be a lower limit to the true width.
The expected width of the line is $2 v_{rot}$ sin$i$ if it is broadened by
galactic rotation, where $i$ is the inclination angle. This implies
a projected rotation velocity along the quasar sightline of $>180$
km s$^{-1}$, which is consistent with earlier estimates.

   If indeed the absorbing galaxy is a rotating disk, one can estimate
the total dynamic mass of the galaxy. As argued above, the rotation
velocity of the galaxy is probably 200 km s$^{-1}$ or more.
The corresponding
dynamic mass is then $v_{rot}^2R/G\geq1.6\times 10^{11} M_{\odot}$
for $R\simeq 17$ kpc, consistent with the Djorgovski et al estimate.

\subsection{Other Properties of the Absorbing Galaxy}

   Is the DLA absorber at $z_{damp}=3.15$ toward Q 2233+1310
unusual as compared to other DLA absorbers?
Other than the fact that it is one of the two DLA systems 
for which Ly$\alpha$ emission from the absorbers has been reported (see 
next section for the discussion of the only other case), we are not aware
of any peculiarity about this system. Based on a preliminary analysis of
the metal absorption lines, we estimate that this DLA galaxy
has a metallicity [Fe/H]$=-1.4\pm0.1$, very similar to other DLA 
galaxies between $2<z<3$ (cf, Pettini et al 1994; Lu et al 1996). 
The relative abundances
of N, O, Al, Si, S, and Fe also appear similar to those studied by 
Lu et al (1996). In all respects, this appears to be just an 
average DLA system.

  The comparison of the absorption profiles between the low ionization 
species and the high ionization species is particularly informative.
Over the velocity range $-300<v<-160$ km s$^{-1}$ where the strongest
low ionization absorption occurs, the high ion absorption lines are
very weak, indicating that the gas at these velocities must be substantially
neutral.
The strongest C IV and Si IV absorption occurs at $v\simeq -120$ km s$^{-1}$,
where the low ion absorption is essentially absent. In fact, one can almost
 make the case that the high ion absorption appears to avoid the low ion
absorption in terms of the positions of the absorption components (figure 1). 
In general, the high ion absorption line
profiles show no similarity to those of the low ionization species in
DLA galaxies (Wolfe et al 1994; Prochaska \& Wolfe 1996; Lu et al 1996), 
suggesting that the bulk of high ion gas
may arise from regions physically unrelated to the low ion gas.
The Q 2233+1310 DLA system evidently fits in that pattern. Because the
high ion absorption extends to as large negative velocity as the low ion
lines, it is possible
that the weak high ion absorption at $v<-160$ km s$^{-1}$ arises
from the disk (although probably not co-spatial with the bulk of low
ion gas in the disk) so they are also spread out
 by galactic rotation. The strong
high ion absorption component at $v\simeq -120$ km s$^{-1}$ is possibly 
associated with a halo cloud outside the disk.

\subsection{The $z_{damp}=2.81$ System toward Q 0528$-$2505}
 
    The technique used here to infer the kinematics of the absorbing
galaxy is applicable only when the systemic redshift of the absorbing
galaxy is known. The only other case reported in the literature of
the detection of Ly$\alpha$ emission from a DLA system 
is by Warren \& Moller (1996) for the $z_{damp}=2.81$ system toward
Q 0528$-$2505. In this case, three Ly$\alpha$ emitters (dubbed S1,
S2, and S3 by the original authors) are reported
with $z=$2.8136, 2.8097, and 2.8126 and with separations from the quasar
line of sight 1.22" (9 kpc), 11.8" (88 kpc), and 20.6" (154 kpc),
respectively, for $q_0=0.1$ and $H_0=75$.
Warren \& Moller argued that S1 is the most likely candidate for
the actual absorber because of its closeness to the quasar sightline. 

Figure 2 shows the absorption profiles of selected
metal lines in the system (detailed analysis of this DLA system
was  given in Lu et al 1996). The velocities
are relative to the rest frame of S1. The velocities corresponding to the
redshifts of S2 and S3 are also indicated. In this case, no clear
signature of galactic rotation is noticeable in the absorption
line profiles. The absorption profiles
(e.g., S II$\lambda$1253) appear to show edge-leading asymmetry,
but the velocity of S1 does not line up with either edge of the profile.
In addition, the velocity spread of the absorption components is 
so large ($\geq 300$ km s$^{-1}$) that it seems 
unlikely that a single galaxy is responsible for the absorption.
The redshift of the DLA absorption ($z_{damp}=2.81$)
is, in fact, slightly higher than the emission redshift of the background 
quasar itself ($z_{em}=2.78$), which makes it likely that the
absorbing galaxy is within the same cluster of galaxies as the quasar
itself\footnote{Q 0528$-$2505 is a radio-loud quasar, and there is strong
evidence that radio-loud quasars tend to live in rich clusters of 
galaxies (cf. Yee \& Green 1987).}. The three individual Ly$\alpha$
emitters may then plausibly be interpreted as either individual galaxies
in the cluster center or subgalactic fragments that are still in the process
of assembly (Warren \& Moller 1996). The large velocity spread of the
metal lines  may be explained if the sightline samples 
more than one galaxy or subgalactic fragment.

\section{DISCUSSION}

By comparing the observed metal absorption line profiles with simulated
model profiles, Prochaska \& Wolfe (1997) find strong statistical evidence
that DLA absorbers at $<z>\sim2.5$ are rotating disks
with a most likely circular velocity $v_{rot}\simeq 250$ km s$^{-1}$.
The technique discussed here presents a way to test
independently the hypothesis that DLA absorbers are young 
disk galaxies. The analysis of Prochaska \& Wolfe (1997) 
has the virtue of a large sample
size, but the results are only statistical. The technique presented here
allows for the test of the galactic rotation hypothesis for individual
cases, but is only applicable when the systemic redshift of the absorber
is known. The two techniques should complement each
other well in revealing the nature of DLA  absorbers
at $z>2$, when it becomes extremely difficult to determine 
the morphology of the absorbing galaxies using standard imaging techniques.
The combination of rotation velocities as inferred from absorption line
analysis with information on the impact parameters 
found from imaging studies allows
for the estimate of the typical mass of the absorbing galaxies.
Such results should provide significant constraints on theories of structure
formation in the early universe.

\acknowledgements
 The authors are grateful to Art Wolfe and Jason X. Prochaska  for
communicating their results on the galactic rotation analysis
of damped Ly$\alpha$ systems in advance of publication, and for
stimulating discussion on the subject. We also thank George Djorgovski
and Chuck Steidel for useful comments on the manuscript.
 This work was based on observations obtained at the
W. M. Keck Observatory, which is operated jointly 
by the California Institute of Technology and the University
of California. We thank the observatory staff for assisting with 
the observations. LL acknowledges support from NASA through grant number 
HF1062.01-94A.  WWS was supported by NSF grant AST95-29073.

\begin{figure}
\plotone{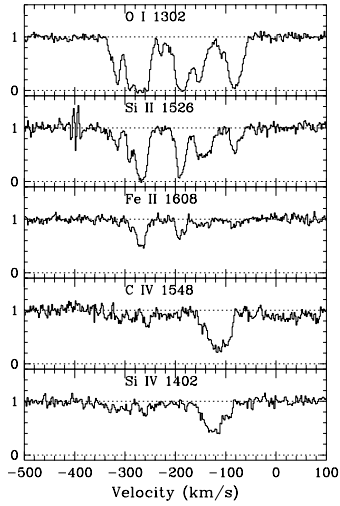}
\caption{Normalized profiles of selected absorption lines in the
$z_{damp}=3.15$ damped Ly$\alpha$ system toward Q 2233+1310. 
The velocity is relative to the
rest-frame of the absorbing galaxy, taken to be at $z=3.15300$.
The spectra were obtained with the Keck telescope and have a
resolution of FWHM=6.6 km s$^{-1}$.}
\end{figure}

\begin{figure}
\plotone{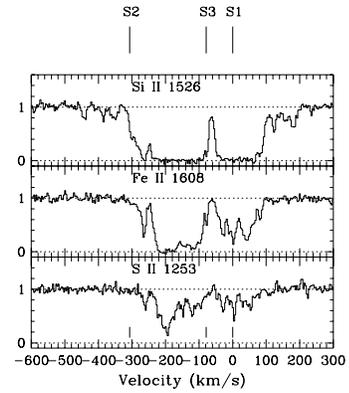}
\caption{Normalized profiles of selected absorption lines in the
$z_{damp}=2.81$ damped Ly$\alpha$ system toward Q 0528$-$2505.
 The velocity is relative to the
rest-frame of S1 (see text), taken to be at $z=2.81360$.
The spectra were obtained with the Keck telescope and have a
resolution of FWHM=6.6 km s$^{-1}$.}
\end{figure}


\begin{references}

\reference{}
Djorgovski, S.G., Pahre, M.A., Bechtold, J., \& Elston, R. 1996, Nature,
  382, 234

\reference{}
Grosbol, P.J. 1985, A\&AS, 60, 261

\reference{}
Kent, S.M. 1985, ApJS, 59, 115

\reference{}
Lanzetta, K.M., \& Bowen D. 1992, ApJ, 391, 48

\reference{}
Lu, L., Sargent, W.L.W., \& Barlow, T.A., Churchill, C.W., \& Vogt, S. 
1996, ApJS, in press

\reference{}
Lu, L., Wolfe, A.M., Turnshek, D.A., \& Lanzetta, K.M. 1993, ApJS, 84, 1

\reference{}
Pettini, M., Smith, L.J., Hunstead, R.W., \& King, D.L. 1994, ApJ, 426, 79

\reference{}
Prochaska, J.X., \& Wolfe, A.M. 1996, ApJ, 470, 403

\reference{}
Prochaska, J.X., \& Wolfe, A.M. 1997, ApJ, submitted

\reference{}
Warren, S.J.,  \& Moller, P. 1996, A\&A, 311, 25

\reference{}
Wolfe, A.M. 1995, in QSO Absorption Lines, ed. G.Meylan (Springer-Verlag), 13

\reference{}
Wolfe, A.M., et al, 1994, ApJ, 435, L101

\reference{}
Wolfe, A.M., Turnshek, D.A., Smith, H.E., \& Cohen, R.D. 1986, ApJS, 61, 249

\reference{}
Yee, H.C., \& Green, R.F. 1987, ApJ, 319, 28

\end{references}
\end{document}